\documentclass[letterpaper, 10 pt, conference]{ieeeconf}  % Comment this line out
\usepackage{mathtools}
\IEEEoverridecommandlockouts                              % This command is only
\usepackage{amsmath}
\usepackage{amssymb}
\usepackage{calrsfs}
\DeclareMathAlphabet{\pazocal}{OMS}{zplm}{m}{n}
\usepackage{xcolor}
\usepackage{xfrac}
\usepackage{dsfont}
\usepackage{siunitx}
\usepackage{algorithm}
\usepackage{pgfplots} %added
\usepackage{algpseudocode}
\usepackage{tikz}
\usepackage{mathabx}
\usepackage{booktabs}
\usepackage{multirow}
\usepackage{threeparttable} % Package for table notes
\usepackage{graphicx}
\usepackage{pgf}  % 
\usepackage{pgfplots}
\usepackage{caption}  % Optional: Better control of captions

\usepackage{url}

\DeclareRobustCommand{\uppartial}{\text{\rotatebox[origin=t]{20}{\scalebox{0.95}[1]{$\partial$}}}\hspace{-1pt}}

\usepackage{setspace}
\title{\LARGE \bf
Learning to Solve Parametric Mixed-Integer Optimal Control Problems via Differentiable Predictive Control
}

\author{Ján Boldocký\authorrefmark{1}, Shahriar Dadras Javan\authorrefmark{2}, Martin Gulan\authorrefmark{1}, Martin Mönnigmann\authorrefmark{2}, Ján Drgoňa\authorrefmark{3}% <-this % stops a space
\thanks{This work was supported by the European Union\textquotesingle{s} Horizon Europe under the grant no. 101079342 and by the Slovak Research and Development Agency under the grants APVV-18-0023 and APVV-22-0436. Ján Boldocký thanks for financial contribution from the STU Grant Schemes for Support of Young Researchers and Excellent Teams of Young Researchers.}%(Fostering Opportunities Towards Slovak Excel-lence in Advanced Control for Smart Industries)
\thanks{\authorrefmark{1}
Ján Boldocký and Martin Gulan are with the Faculty of Mechanical En-gineering, Slovak University of Technology in Bratislava, Slovakia
        (email: {\tt\footnotesize \{jan.boldocky,martin.gulan\}@stuba.sk}).}%
\thanks{\authorrefmark{2}Shahriar Dadras Javan and Martin Mönnigmann are with the Automatic Control and Systems Theory, Ruhr University Bochum, Germany (email: {\tt\footnotesize \{shahriar.dadrasjavan,martin.moennigmann\}@rub.de}).}%
\thanks{\authorrefmark{3}Ján Drgoňa is with the Department of Civil and Systems Engineering, Johns Hopkins University, MD, USA 
(email: {\tt\footnotesize jdrgona1@jh.edu}).}%
}

\begin{document}

\maketitle
\thispagestyle{empty}
\pagestyle{empty}

%%%%%%%%%%%%%%%%%%%%%%%%%%%%%%%%%%%%%%%%%%%%%%%%%%%%%%%%%%%%%%%%%%%%%%%%%%%%%%%%
\begin{abstract}

% We introduce a novel approach to solving input- and state-constrained mixed-integer optimal control problems using Differentiable Predictive Control (DPC). The proposed approach follows a differentiable programming paradigm in which we establish an explicit neural control policy that maps control parameters to integer- and continuous-valued decision variables by differentiating the quadratic model predictive control objective function through the closed-loop response of system dynamics, where gradients of integer decision variables are approximated using three different methods. We evaluate our approach on a small-scale thermal energy system, comparing its performance to the optimal solution for different lengths of the prediction horizon in terms of suboptimality and evaluation time. Obtained results indicate that the proposed approach can achieve near-optimal performance while significantly reducing evaluation time.

We propose a novel approach to solving input- and state-constrained parametric mixed-integer optimal control problems using Differentiable Predictive Control (DPC). Our approach follows the differentiable programming paradigm by learning an explicit neural policy that maps control parameters to integer- and continuous-valued decision variables. This policy is optimized via stochastic gradient descent by differentiating the quadratic model predictive control objective through the closed-loop finite-horizon response of the system dynamics. To handle integrality constraints, we incorporate three differentiable rounding strategies. The approach is evaluated on a conceptual thermal energy system, comparing its performance with the optimal solution for different lengths of the prediction horizon. The simulation results indicate that our self-supervised learning approach can achieve near-optimal control performance while significantly reducing inference time by avoiding online optimization, thus implying its potential for embedded deployment even on edge devices.

% We evaluated our approach on a small-scale thermal energy system, comparing its performance with the optimal solution for different lengths of the prediction horizon in terms of suboptimality and solution time. The results indicate that our self-supervised learning approach can achieve near-optimal performance while significantly reducing solution time.
\end{abstract}

%%%%%%%%%%%%%%%%%%%%%%%%%%%%%%%%%%%%%%%%%%%%%%%%%%%%%%%%%%%%%%%%%%%%%%%%%%%%%%%%
\section{INTRODUCTION}

Mixed-integer optimal control problems (MI-OCPs) occur frequently in modern engineering applications across various domains, ranging from energy management and robotics to transportation and industrial process control. This class of problems is defined for systems in which one or more decision variables must take integer values. Hence, solving the OCP requires optimizing both continuous and discrete variables while ensuring feasibility with respect to system dynamics and constraints. 

A widely adopted framework for handling constrained MI-OCPs is mixed-integer model predictive control (MI-MPC) \cite{mcallister2022advances}. Here, the optimal control problem is defined as the mixed-integer program (MIP), which is in the class of NP-hard problems \cite{pia2017}.  In the case of the quadratic norm performance index, affine system dynamics, and linear constraints, the problem can be cast as a mixed-integer quadratic program (MIQP)  \cite{borrelli2017}. 
% However, the optimization problem is further scaled due to the introduction of a prediction horizon.
The inherent combinatorial nature of MIQP may lead to exponential complexity scaling, making real-time optimization particularly challenging. Traditional approaches for solving these problems rely on well-established methods, such as branch and bound or cutting plane, which, despite their theoretical optimality guarantees, may quickly become computationally prohibitive for real-time applications \cite{lin2011fast}. 

In engineering practice, the problem of computational complexity is usually avoided at the expense of introducing some margin of suboptimality. For example, in \cite{frick2015}  the authors propose reducing the computational time by relaxing the minimization of the objective function along the prediction horizon while preserving the integrality constraints and feasibility conditions, whereas in \cite{javan2024}, the authors opted for the approach of omitting the integrality constraints along the prediction horizon while preserving the optimality of the relaxed problem. In \cite{marcucci2021}, a warm start algorithm is proposed, which effectively prunes a portion of the search space for the branch-and-bound algorithm. 
In some cases, the computational cost is reduced with the help of machine learning methods. For example, in \cite{karg2018} and \cite{cauligi2020}, the so-called imitation learning approach is used, where a neural network is trained to mimic optimal trajectories, while in \cite{lohr2020}, \cite{reiter2024}, and \cite{cauligi2022}, the authors use a similar concept to presolve the problem by predicting integer-valued control variables and subsequently solving the remaining convex quadratic program for continuous-valued variables. Furthermore, in \cite{mitrai2024}, the authors proposed machine learning-based surrogate models that approximate Benders cuts, effectively reducing the solution space, however, still requiring a solution to the reduced mixed-integer program.
% Solving large-scale optimal control problems efficiently poses a significant challenge in terms of real-time control.
Recent studies have leveraged reinforcement learning to reduce the solution time of MI-OCPs. For example, in \cite{guo2022}, a Q-learning algorithm is used to obtain the control policy, and more recently, in \cite{xu2025}, a novel algorithm is introduced to solve MI-OCPs, leveraging actor-critic and Q-learning methods to manage continuous and integer-valued action variables simultaneously. 

Moreover, in \cite{dua1999algorithms, dua2002}, a multiparametric programming approach was considered, where the optimal solution map is precomputed offline. This enables real-time control by evaluating the precomputed explicit solution instead of solving the optimization problem online; however, this is limited due to substantial scaling issues. 
Differentiable Predictive Control (DPC)   follows a similar idea, where the parametric OCP is posed as a differentiable program \cite{drgona2022,drgona2024}, and the solution map is represented by a differentiable parametric function optimized via stochastic gradient descent, making it a scalable alternative to the multiparametric approach.

% The main contribution of this work is extending the DPC framework for solving MI-OCPs by 

% In this work, we extend the DPC framework by enabling it to solve MI-OCPs.
In this work, we effectively employ the DPC framework to solve MI-OCPs by imposing the integrality constraints using three different methods, where the gradients of integer decision variables are approximated using differentiable surrogates.
% , where the integrality of integer-valued decision variables is imposed by three distinct differentiable rounding strategies.
% gradients of integer-valued decision variables are approximated using three distinct methods. 
Section \ref{sec:2} presents the nominal formulations of MI-MPC and parametric MI-OCP. Section \ref{sec:3} presents the methodology for the relaxation of integrality constraints using differentiable rounding strategies within the proposed mixed integer differentiable predictive control (MI-DPC) framework. Section \ref{sec:4} discusses the numerical results of the proposed MI-DPC approach in a disturbance rejection simulation of a second-order linear time-invariant system. The results suggest that the proposed framework may provide an effective and scalable approach to obtain a near-optimal explicit solution of parametric MI-OCPs.
% \newline % 4 rows to go
% \newline
% \newline
% \newline

%%%%%%%%%%%%%%%%%%%%%%%%%%%%%%%%%%%%%%%%%%%%%%%%%%%%%%%%%%%%%%%%%%%%%%%%%%%%%%%%
\section{Problem formulation}
\label{sec:2}
In the following, we present the nominal MI-MPC problem and the analogous formulation for DPC, specific to the numerical example presented in Section \ref{sec:4}. 

We assume the control problem of a linear time-invariant discrete-time deterministic system given by
\begin{equation}\label{eq:state-space}
    x_{k+1}=Ax_k+B_\mathrm{u}u_k+B_\delta\delta_k+Ed_k,
\end{equation}
where $u\in\mathbb{R}^{n_u}$ and $\delta\in\mathbb{Z}^{n_\delta}$ denote the continuous- and integer-valued control inputs, respectively, $d\in\mathbb{R}^{n_d}$ denotes the vector of known disturbances and $x\in\mathbb{R}^{n_x}$ is the state vector at discrete time step $k$. The matrices $A$, $B_u$, $B_\delta$, and $E$ are of respective dimensions. Furthermore, we assume that the system is controllable and is subjected to linear input and state constraints
\begin{equation}
    x_k\in\pazocal{X}, \;u_k\in\pazocal{F}, \;\delta_k\in\pazocal{D},\;\forall k\geq0,
\end{equation}
where $\pazocal{X}\subset\mathbb{R}^{n_x}$ and $\pazocal{F}\subset\mathbb{R}^{n_u}$  are compact convex sets. The constraints contained in the set of feasible integers $\pazocal{D}\subset\mathbb{Z}^{n_\delta}$ are commonly referred to as integrality constraints.

\subsection{Mixed-Integer Model Predictive Control}
For simplicity, we consider the nominal single-shooting MI-MPC formulation, with the control objective defined by stage cost $\ell_\pazocal{K}$ as the squared Euclidean weighted norm of the control error, integer-, and continuous-valued control inputs and a terminal cost $\ell_\pazocal{N}$. The control error is defined as the difference between the state $x$ and the reference $r\in\mathbb{R}^{n_x}$. We define stage and terminal costs as
\begin{subequations}
    \begin{align}
        \ell_\pazocal{K}(x_k,r_k,u_k,\delta_k)&=
       \!\|x_k\!-\!r_k\|_Q^2+\|u_k\|_R^2+\|\delta_k\|^2_\rho,\\
       \ell_\pazocal{N}(x_k,r_k)&=\|x_k\!-\!r_k\|_P^2,
    \end{align}
\end{subequations}
% where the weight matrices $P\succeq0$, $Q\succeq0$, $R\succ0$, $\rho\succ0$ are of obvious dimensions.
where the weight matrices $P$, $Q$, and $R$, $\rho$ are positive semi-definite and positive definite, respectively, and of obvious dimensions. 
Next, using these control objective components, let us formulate the MI-MPC problem as
\begin{subequations} \label{eq:mi-mpc}
    \begin{align}
       \min_{u,\delta}&\;\ell_\pazocal{N}(x_N,r_N)
       \!+\!\sum_{k=0}^{N-1}
       \ell_\pazocal{K}(x_k,r_k,u_k,\delta_k) \label{eq:mi-mpc-loss} \\
         \text{s.t.}& \;
        x_{k+1}  = Ax_k+B_\mathrm{u}u_k+B_\delta\delta_k+Ed_k, \label{eq:mi-mpc-model} \\
       &\quad u_k  \in \pazocal{F},\;\;\delta_k \in \pazocal{D}, \quad \forall k\in\{0,1,\dots,N\!-\!1\},\label{eq:mi-mpc-input-constraints}
       \\
        &\quad x_k  \in \pazocal{X}, \quad\quad\quad\quad\;\;\; \forall k\in\{0,1,\dots,N\}, \label{eq:mi-mpc-state-constraints}
    \end{align}
\end{subequations}
where $N\in\mathbb{Z}^+$ denotes the length of the prediction horizon.

Now, after substituting the system dynamics \eqref{eq:mi-mpc-model} into the loss function \eqref{eq:mi-mpc-loss} and transforming the constraints \eqref{eq:mi-mpc-input-constraints} and \eqref{eq:mi-mpc-state-constraints} accordingly, the original MI-MPC problem \eqref{eq:mi-mpc} can be cast as a mixed-integer quadratic program:
%\vspace{-1.5ex}
\begin{subequations} \label{eq:miqp}
    \begin{align}
       \min_{z_k}&\;z_k^\mathsf{T}\pazocal{H}_k z_k - \pazocal{G}_k^\mathsf{T}z_k\\
       \text{s.t.}&\;
       \Omega z_k\leq\omega_k,
    \end{align}
\end{subequations}
where
$z_k\!=\!\big[\upsilon_k^\mathsf{T},\upsilon_{k+1}^\mathsf{T}, \dots, \upsilon_{k+N-1}^\mathsf{T}\big]^\mathsf{T}\in\mathbb{R}^{N(n_u+n_\delta)}$ combines the integer- and continuous-valued control inputs $\upsilon_k\!=\!\big[u_k, \delta_k\big]^\mathsf{T}$ for the entire horizon length. 
% For a detailed mathematical derivation, we refer reader to \cite{maciejowski2007predictive}.

As mentioned above, due to the combinatorial nature, these problems are computationally demanding to solve, where finding the optimal solution may scale exponentially with the size of the problem. The goal of this paper is to avoid the computational burden of solving the given problem by formulating it within the DPC framework, which was shown to be computationally efficient in solving medium to large-scale MPC problems \cite{DRGONA2021buildings}.

\subsection{Parametric Mixed-Integer Optimal Control Problem}
 Considering the system \eqref{eq:state-space}, we formulate the parametric MI-OCP as
\begin{subequations} \label{eq:dpc}
    \begin{align}
       \min_{\theta}  \, \!\mathbb{E}_{  \xi_k \sim P_{ \xi} }\,&\!  
       \bigg[\ell_\pazocal{N}(x_N, r_N)+\sum_{k=0}^{N-1}\!\!   \ell_\pazocal{K}(x_k, r_k, u_k, \delta_k)\bigg]  \label{eq:dpcloss} \\
       \text{s.t.} \;
        x_{k+1} & = Ax_k+B_\mathrm{u}u_k+B_\delta\delta_k+Ed_k, \label{eq:dpc-model} \\
        [u_k,\;\delta_k] & = {\pi}_{\theta}(\xi_k), \label{eq:dpcinput} \\
         u_k  \in \pazocal{F}&,\;\delta_k \in \pazocal{D}, \quad \forall k\in\{0,1,\dots,N\!-\!1\}, \label{eq:dpcinputconstraint} \\
        x_k  \in \pazocal{X}&, \quad\quad\quad\quad\;\; \forall k\in\{0,1,\dots,N\},\label{eq:dpcstateconstraint}
        % \boldsymbol{\xi}_k & \in \Xi^{\mathrm{ctrl}} \subset \mathbb{R}^n,
    \end{align}
\end{subequations}
where \eqref{eq:dpcloss} represents a parametrized MPC objective function, sampled with control parameters $\xi_k$ of a known probability distribution $P_\xi$. The vector of control parameters consists of all known or estimated variables that are used to calculate the control input. For example, in the case of a constant reference $r$, constant state and input constraints, and known future disturbances $d$, the vector would be defined by initial condition $x_k$ and a sequence of future disturbances as $\xi_k\!=\![x_k^\mathsf{T},\,d_k^\mathsf{T},\dots,d_{k+N-1}^\mathsf{T}] \in \mathbb{R}^{n_\xi}$. These conditions hold for the numerical example presented in Section \ref{sec:4}; henceforth $\xi_k$ will be assumed as such. Here, the system dynamics model \eqref{eq:dpc-model}, is used to obtain the $N$-steps long closed-loop trajectories, where the control inputs $u_k$ and $\delta_k$ are calculated by the control policy $\pi_\theta\!:\! \mathbb{R}^{n_\xi}\!\mapsto\!\mathbb{R}^{n_u}\!\times \!\mathbb{R}^{n_\delta}$, which is parametrized by a set of parameters $\theta$.

\section{Methodology}
\label{sec:3}

\subsection{Differentiable Predictive Control}
Differentiable predictive control (DPC) is an offline policy optimization method in which the parametric optimal control problem is formulated as a differentiable program~\cite{drgona2022,drgona2024}. Here, the control policy can be represented by an arbitrary differentiable function, most commonly a neural network, for its parameter efficiency and expressive capacity.
To be able to cast MI-OCP \eqref{eq:dpc} as a differentiable program, in DPC, we consider the relaxation of state and input constraints using the penalty, Lagrangian relaxation, or Barrier methods. This gives us a dual problem that is convex, assuming that the primal problem is convex \cite{boyd2004}.
With the penalty method-based relaxation, the problem is formulated as
\begin{subequations} \label{eq:dpc-dual-hard}
    \begin{align}
        \begin{split}
        \begin{aligned}[t]
       \min_{\theta}  \, \mathbb{E}_{\xi_k \sim P_{ \xi} }&\,  
       \bigg[\|x_N\!-\!r_N\|_P^2
       \!+\!\sum_{k=0}^{N-1}\!\|x_k\!-\!r_k\|_Q^2\\ +\|u_k\|_R^2&+\|\delta_k\|^2_\rho\!+\!c_xq(x_k, x_N)\!+\!c_up(u_k)\bigg]   
       \end{aligned}
       \end{split}
       \label{eq:dpc-dual-hard-loss} \\
       &\text{s.t.} \;
        x_{k+1}=Ax_k+B_\mathrm{u}u_k+B_\delta\delta_k+Ed_k, \label{eq:dpc-model-dual-hard}\\ 
       & \;[u_k,\;\delta_k] = {\pi}_{\theta}(\xi_k).& \label{eq:dpc-policy-hard}
    \end{align}
\end{subequations}

Assuming differentiability of the model \eqref{eq:dpc-model-dual-hard} and the con-trol policy \eqref{eq:dpc-policy-hard}, the parametric OCP \eqref{eq:dpc-dual-hard} can be implemented with a differentiable programming framework, such as PyTorch or Julia, where parameters of the control policy $\theta$ are optimized using some variant of stochastic gradient descent, so that the loss function \eqref{eq:dpc-dual-hard-loss} is minimized. To compute the gradients of the total loss $\ell$ with respect to the policy parameters, we employ the chain rule as
\begin{align}
\label{eq:grad}
    \nabla_{\theta} \ell = & \, \frac{ \uppartial\ell_\pazocal{K}}{\uppartial \theta}  +  
    \frac{ \uppartial \ell_\pazocal{N}}{\uppartial \theta} +
    \frac{ \uppartial q(x)}{\uppartial \theta} + 
    \frac{ \uppartial p(u)}{\uppartial \theta} \nonumber\\
     = & \, \frac{ \uppartial\ell_\pazocal{K}}{\uppartial x} \frac{ \uppartial x}{\uppartial u} \frac{ \uppartial u}{\uppartial \theta}
     +\frac{ \uppartial\ell_\pazocal{K}}{\uppartial x} \frac{ \uppartial x}{\uppartial \delta} \frac{ \uppartial \delta}{\uppartial \theta}
     +\frac{ \uppartial\ell_\pazocal{K}}{\uppartial u}  \frac{ \uppartial u}{\uppartial \theta} +\frac{ \uppartial\ell_\pazocal{K}}{\uppartial \delta}  \frac{ \uppartial \delta}{\uppartial \theta}\nonumber\\
    %  &+ \frac{ \uppartial q(x)}{\uppartial x}  \frac{ \uppartial x}{\uppartial \upsilon} \frac{ \uppartial \upsilon}{\uppartial \theta}
    % + \frac{ \uppartial p(u)}{\uppartial u} \frac{ \uppartial u}{\uppartial \theta} 
    &+ \frac{ \uppartial \ell_\pazocal{N}}{\uppartial x} \frac{ \uppartial x}{\uppartial u} \frac{ \uppartial u}{\uppartial \theta} + \frac{ \uppartial \ell_\pazocal{N}}{\uppartial x} \frac{ \uppartial x}{\uppartial \delta} \frac{ \uppartial \delta}{\uppartial \theta} + \frac{ \uppartial q(x)}{\uppartial x}  \frac{ \uppartial x}{\uppartial u} \frac{ \uppartial u}{\uppartial \theta}\nonumber\\
    &+ \frac{ \uppartial q(x)}{\uppartial x}  \frac{ \uppartial x}{\uppartial \delta} \frac{ \uppartial \delta}{\uppartial \theta}+ \frac{ \uppartial p(u)}{\uppartial u} \frac{ \uppartial u}{\uppartial \theta},
\end{align}
where $\sfrac{ \uppartial u}{\uppartial \theta}$ and $\sfrac{ \uppartial \delta}{\uppartial \theta}$ represent the gradients of continuous and discrete control inputs, respectively, computed using the backpropagation through time algorithm \cite{puskorius1994}.

%%%%%%%%%%%%%%%%%%%%%%%%%%%%%%%%%%%%%%%%%%%%%%%%%%%%%%%%%%%%%%%%%%%%%%%%%%%%%%%%

\subsection{Differentiable Mixed-Integer Neural Control Policy}

The main issue with applying the DPC policy gradients~\eqref{eq:grad} to the MI-OCP problem is the non-differentiability of the mixed-integer control policy.
In this study, we propose a novel relaxation of the integrality constraints~\eqref{eq:dpcinputconstraint} using differentiable nonlinear rounding strategies as a structural part of the control policy. 
In Sections \ref{sec:3a} to \ref{sec:3c}, we provide insight into relaxing the integrality constraints through three distinct rounding strategies, where the gradients $\frac{ \uppartial \delta}{\uppartial \theta}$ are approximated via differentiable surrogates.

To preserve the integrality of the integer-valued control actions computed by the neural control policy $\pi_\theta$, we introduce discrete functions at the output layer. To avoid the problem of non-differentiability of discrete functions, we employ a widely used heuristic paradigm of gradient approximation, originally introduced in \cite{bengio2013} as the Straight-Through Estimator (STE). 
The main idea here is to use a differentiable surrogate of the discrete non-differentiable function to calculate the gradients, enabling approximate gradient-based optimization.

We formulate the control policy $\pi_\theta$ using a feed-forward neural structure composed of three components as
\begin{subequations}
    \begin{align}
        h_k &= f_{\theta^{(h)}}^{(h)}(\xi_k),\label{eq:nn1}\\
        y^{(u)}_k&= f_{\theta^{(u)}}^{(u)}(h_k), \label{eq:nn2}\\
        y^{(\delta)}_k&= f_{\theta^{(\delta)}}^{(\delta)}(h_k),\label{eq:nn3}
        % &a^{(0|\beta)}_k=W^\prime\xi_k+b^\prime,  \;\;\quad \beta\in\{u,\delta\},\\
        % &a^{(l|\beta)}_k=f_a(W^{(l-1|\beta)}a^{(l-1|\beta)}_k+b^{(l-1|\beta)}), \;\; l \in \mathbb{N}_1^L,\\
        % &y_k^{(\beta)} = W^{(L|\beta)} a^{(L|\beta)}_k+b^{(L|\beta)},\label{eq:network_outputs}
    \end{align}
\end{subequations}
where the component $f^{(h)}_{\theta^{(h)}}: \mathbb{R}^{n_{\xi}} \mapsto \mathbb{R}^{n_h}$ projects the control parameters $\xi$ onto a higher-dimensional latent space. The vector of the lifted control parameters $h\in\mathbb{R}^{n_h}$ is then fed into the two neural modules $f^{(\delta)}_{\theta^{(\delta)}}$ and $f^{(u)}_{\theta^{(u)}}$ that output relaxed solutions of integer- and continuous-valued control inputs, $y_k^{(\delta)}$ and $y_k^{(u)}$, respectively.
The parameters of the control policy, $\theta\!=\!\{\theta^{(h)},\theta^{(u)},\theta^{(\delta)}\}$, consist of the parameters of each neural component.  
% The parameters of each module compose the parameters of the control policy $\theta\!=\!\{\theta^{(h)},\theta^{(u)},\theta^{(\delta)}\}$.
In the following, we describe how the integrality constraints on $y^{(\delta)}$ are enforced using different discrete methods and how gradients are approximated using differentiable surrogates.

% In the following, we describe the enforcement of the integrality constraints on $y^{(\delta)}$ using different discrete methods and approximating the gradients using differentiable surrogates.

% where $W^{(l|\beta)}$ and $b^{(l|\beta)}$ denote the weights and biases of a layer $l$ in module $\beta$, respectively, $W^\prime$ and $b^\prime$ denote the parameters of a common input layer, $f_a$ is an arbitrary activation function, and $L$ denotes the depth of the two branches. The output vectors of the two branches are denoted by $y^{(\beta)}$, where $y^{(u)}\in \mathbb{R}^{n_u}$ represents the solution of a continuous variable branch and $y^{(\delta)}\in\mathbb{R}$ a relaxed solution of integer variable. In the following, we describe the enforcement of the integrality constraints on $y^{(\delta)}$ with differentiable rounding strategies using STE with Sigmoid and softmax functions.

\subsection{Straight-Through Estimator with Sigmoid}\label{sec:3a}
If the integrality constraint~\eqref{eq:dpcinputconstraint} is enforced by a rounding operation $ \delta_k = \lfloor y_k^{(\delta)}\rceil$, the Sigmoid function may be used as a differentiable surrogate. In the STE setting, the Sigmoid, $\sigma(x)\!=\!(1+e^{-x})^{-1}$, can be parameterized by a slope coefficient, $\eta\!\in\!\mathbb{R}_{>1}$, where a larger $\eta$ results in a more accurate approximation of the non-smooth rounding function, however, at the expense of a steeper gradient. With this strategy, we transform the relaxed solutions $y^{(\delta)}\in\mathbb{R}^{n_\delta}$ and $y^{(u)}\in\mathbb{R}^{n_u}$ to control inputs $\delta$ and $u$ as 
\begin{subequations}
    \begin{align}
        \delta_k &= \lfloor y_k^{(\delta)}\rceil, \label{eq:rounding-nearest} \\
        u_k &= y^{(u)}_k,
    \end{align}
\end{subequations}
where integrality is enforced by rounding the network output $y^{(\delta)}$ to the nearest integer in \eqref{eq:rounding-nearest}. Here, the gradient of the integer variable $\delta_k$ is approximated by differentiating the Sigmoid surrogate given as
% \begin{subequations}
    \begin{align}
       \nabla\delta_k &\approx \nabla\sigma\big(\eta (y_k^{(\delta)}-\delta_k-t)\big)\nonumber\\
        \approx\sigma&\big(\eta(y_k^{(\delta)}-\delta_k-t)\big)\!\Big(1-\sigma\big(\eta(y_k^{(\delta)}-\delta_k-t)\big)\!\Big) \eta, \label{eq:sigmoid-grad}
    \end{align}
% \end{subequations}
where $t\!=\!0.5$ is the rounding threshold.
However, this strategy is applicable only when feasible integers are evenly spaced, such as $\{0,1,2\}$. If integers are not evenly spaced,
e.g. $\{0,1,5\}$,
a categorical approach may be leveraged. 

\subsection{Straight-Through Estimator with Softmax}\label{sec:3b}
Relaxing the integrality constraints can be addressed by formulating the problem as a categorical classification. Here, the relaxed integer solution $y_k^{(\delta)}$ constitutes a set of feasible integer probability score vectors $S$ per $j$-th integer control input at time $k$ as
% \begin{subequations}
\begin{equation}
\begin{aligned}
% y^{(\delta)}_k &= \bigcup_{j=1}^{n_\delta}\ S_{k|j},\quad\\
% y^{(\delta)}_k &=  \{S_{k|j}, \dots,S_{k|n_\delta}\}, \quad j\in\{1,\dots,n_\delta\},\\
&y^{(\delta)}_k =  \{S_{k|j}\}_{j=1}^{n_\delta},\\
& \phantom{ = } \text{with}\; S_{k|j} = [s_{k|1},\dots,s_{{k|L_j}}]^\mathsf{T}, \forall j, L_j\geq2,
\end{aligned}
\end{equation}
% \end{subequations}
where $s\in\mathbb{R}$ denotes the probability score of a correspond-ing feasible integer of index $i\!\in\!\{1, \dots, L_j\}$. Naturally, every integer control input $\delta$ can attain $L_j\!\geq\!2$ integer values. % For example, if the controlled system has one integer control input with four feasible integer values $\{0,1,2,3\}$, the dimension of relaxed integer solution would be $y^{(\delta)}\in \mathbb{R}^{1\times4}$. 
These scores are transformed into normalized probabilities $\hat{s}$ using a Softmax function. Moreover, as introduced in \cite{jang2017}, we consider perturbing the probabilities by introducing the noise sampled from the Gumbel distribution, which promotes further exploration of the search space during training.  The normalized probabilities are calculated as
\begin{subequations}
\begin{align}
            \hat{s}_{k|i} &= \frac{e^{\big((\log(s_{k|i})+g_{k|i})\tau^{-1}\big)}}{\sum_{m=1}^{L_j}e^{ \big((\log(s_{k|m})+g_{k|m})\tau^{-1}\big)}}, \;\; \forall i\in\{1,\dots,L_j\},\\
            \hat{S}_{k|j}&=[\hat{s}_{k|1},\dots,\hat{s}_{{k|L_j}}]^\mathsf{T}, \quad \hat{s}_{k|i}\in (0, 1),
\end{align}
\end{subequations}
where  $g$ denotes noise samples drawn from  $\text{Gumbel}(0,1)$ distribution and $\tau\!\in\!\mathbb{R}_{>0}$ denotes the temperature coefficient, which polarizes the resulting probability vector $\hat{S}$ if $\tau\!<\!1$. For probability scores it holds that $\sum_m \hat{s}_{k|m}\!=\!1$. Next, we can calculate arbitrary integer control inputs by introducing the set of feasible integer vectors $\pazocal{A}\!=\!\{A_{j}\}_{j=1}^{n_\delta}$, where $A_j\!=\![a_1,\dots,a_{L_j}]^\mathsf{T}$, and one-hot encoded probability vectors $\bar{S}_{k|j}\!=\![\bar{s}_{k|1},\dots,\bar{s}_{k|L_j}]^\mathsf{T}$ as
\begin{subequations}
    \begin{align}
        % asdf
        \bar{s}_{k|i} &=  \begin{cases} \label{eq:one-hot}
            1, & \text{if } i = \arg\underset{m}{\max}\;\hat{s}_m\; \text{for each }m,\\
            0, & \text{otherwise}, 
        \end{cases} \\ 
        \delta_{k|j} &= 
        % \big[\bar{s}_{k|1},\dots,\bar{s}_{k|L_j}\big] \cdot 
        % \big[a_{1},\dots,a_{L_j}\big]^\mathsf{T},\\
        \bar{S}_{k|j}^\mathsf{T} A_j, \label{eq:onehot-matmul}\\
        u_k &= y^{(u)}_k.
    \end{align}
\end{subequations}
In \eqref{eq:onehot-matmul}, the binary one-hot encoded probability vector $\bar{S}$ ensures that one of the feasible integers $a_i$ is chosen to represent the integer control variable, while the other integers are multiplied by zero value. Since one-hot encoding \eqref{eq:one-hot} is not a differentiable function, the gradient of integer control variable $\nabla\delta$ is calculated using the vector of polarized pro-babilities $\hat{S}$ as
\begin{equation}
    \nabla\delta_{k|j} \approx \nabla\big(\hat{S}_{k|j}^\mathsf{T} A_j\big),
\end{equation}
where as the temperature coefficient decreases $(\tau\!\to\!0)$, the polarization of the probability vector becomes more pronounced, leading to a more accurate gradient approximation, however, at the expense of numerical stability.

\subsection{Learnable Threshold Approach}\label{sec:3c}
The Learnable Threshold (LT) approach has recently been introduced in \cite{tang2025}, where the authors propose a self-supervised learning framework to solve mixed-integer nonlinear programs. The basic idea of the LT method is to predict a continuously relaxed solution $y^{(\delta)}_k\!=\!\phi_{1|\theta_1}(\xi_k)$ using a neural model $\phi_{1|\theta_1}
% :\mathbb{R}^{n_\xi}\mapsto \mathbb{R}^{n_u+n_\delta}
$, and compute a correction of the relaxed solution $q_k\!=\!\phi_{2|\theta_2}(\xi_k,y_k)$ using second neural model $\phi_{2|\theta_2}
% : \mathbb{R}^{n_\xi}\times\mathbb{R}^{n_u+n_\delta} \mapsto \mathbb{R}^{n_u}\times\mathbb{Z}^{n_\delta}
$  as $y_k^{(\delta)}\!\leftarrow\!y^{(\delta)}_k + q_k$.  Subsequently, the rounding thresholds are calculated $t_k\!=\!\sigma(h^{(\delta)}_k)$ using Sigmoid. Finally, the integer variable is obtained by rounding the continuous solution down $\delta_k\!=\!\lfloor y^{(\delta)}_k \rfloor$ if its fractional value is less than the computed threshold $(y^{(\delta)}_k\!-\!\lfloor y^{(\delta)}_k\rfloor\!<\!t_k)$, and rounded up $\delta_k\!=\!\lceil y^{(\delta)}_k \rceil$ if its fractional value exceeds the threshold $(y^{(\delta)}_k\!-\!\lfloor y^{(\delta)}_k\rfloor\!>\!t_k)$. The gradient of the rounding operation is approximated using Sigmoid-based STE, similarly as in \eqref{eq:sigmoid-grad} with an extension of the threshold values as learnable parameters. For further details, we refer the reader to \cite{tang2025}.
%%%%%%%%%%%%%%%%%%%%%%%%%%%%%%%%%%%%%%%%%%%%%%%%%%%%%%%%%%%%%%%%%%%%%%%%%%%%%%%%
\section{NUMERICAL EXAMPLE}
\label{sec:4}
The proposed approach is demonstrated on a linear time-invariant second-order system with two continuous- and one integer-valued control inputs. Additionally, the system states are perturbed by two disturbance variables, which are considered to be known.
The model structure is inspired by \cite{lohr2019mimicking}.
Let us parameterize the system \eqref{eq:state-space} as
% \begin{subequations}
% \begin{align}
\[
\renewcommand{\arraystretch}{0.7}
\begin{aligned}
    A&=\begin{bmatrix}
\alpha_1 & \nu \\
0 & \alpha_2-\nu
\end{bmatrix},\;
B_u=
\begin{bmatrix}
b_1 & 0 \\
0 & b_2
\end{bmatrix},&\; \\
B_\delta&=
\begin{bmatrix}
0  \\
b_3 
\end{bmatrix},\;
E=
\begin{bmatrix}
-b_4 & 0 \\
0 & -b_5
\end{bmatrix},&
\end{aligned}
\]
where the coefficients $\alpha_{1}\!=\!0.9983$ and $\alpha_2\!=\!0.9966$ denote the rates of energy dissipation from thermal storage tanks $x_1$ and $x_2$, respectively, $\nu\!=\!0.001$ denotes the rate of energy transition from tank $x_2$ to tank $x_1$, and coefficients $b_1\!=\!0.075$ and $b_2\!=\!0.075$ denote the thermal efficiency of the two heat pumps $u_1$ and $u_2$, respectively. The discrete control input $\delta_1\!\in\!\{0,1,2,3\}$ represents the number of active heating rods inside the tank $x_2$, each with a power output of \SI{1}{\kilo\watt} and operating with an efficiency of $b_3\!=\!0.0825$. In addition, the coefficients $b_4\!=\!0.0833$ and $b_5\!=\!0.0833$ relate consumption of the energy stored in the tanks $x_1$ and $x_2$ to the disturbance signals $d_1$ and $d_2$, respectively. The parameters given correspond to a system with a sampling period of $T\!=\!300\;\si{\second}$.

Next, we design the parametric OCP \eqref{eq:dpc-dual-hard} in which the ref-erence tracking term is weighted by $Q\!=\text{diag}(1,1)\!$ for both stage and terminal costs equally, i.e. $P\!=\!Q$. The continuous control inputs are penalized by $R\!=\!\text{diag}(0.5,0.5)$ and the discrete input by $\rho\!=\!0.1$. Note that the penalty matrices are not chosen to represent a real-world scenario, but rather to make an adequate benchmark. Specifically, the cost of the heating rods is chosen to be lower than the cost of the heat pump, as we want them to be activated more frequently for demonstrative purposes.
The states of the system are within the limits of $0\!\leq\!x_1\!\leq\!8.4$ and $0\!\leq\!x_2\!\leq\!3.6$, weighted by penalty coefficients of $c_x\!=\!25$ and the continuous control inputs are bounded within $u_1,u_2\!\geq\!0$ and $0\!\leq\!u_1+u_2\!\leq\!8$, with penalty coefficients of $c_u\!=\!25$.
The penalty method was applied for the relaxation of these constraints as defined by~\eqref{eq:dpc-dual-hard}. Finally, the system states $x_1$ and $x_2$ are controlled to constant reference values $r_1\!=\!4.2$ and $r_2\!=\!1.8$, respectively.

\subsection{Controller Synthesis}
\begin{figure*}[tb]
    \begin{center}
        \begingroup
       \footnotesize
       \resizebox{\textwidth}{!}{\input{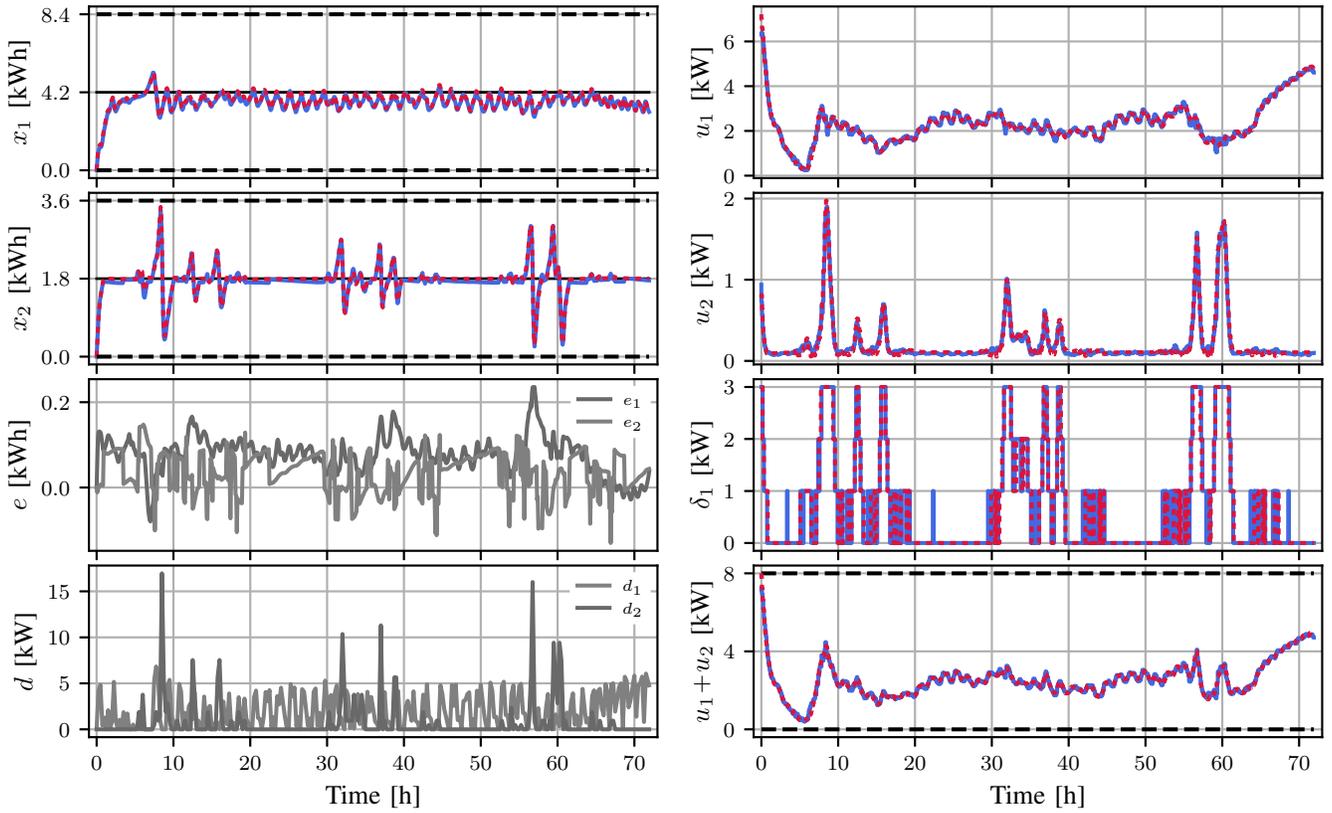}}
    \endgroup
    \end{center}
    \caption{Closed-loop simulation results obtained with MI-DPC with Sigmoid STE (blue), compared with the optimal solution (dashed red), for a horizon of $N\!=20$. The reference states are depicted in black and state/input bounds in dashed black.}
        %Closed-loop simulation results of DPC with Sigmoid STE, compared to the optimal solution for a horizon of $N\!=20$ steps. Note: DPC inputs/states -- blue, reference states -- black, constraints -- dashed black, optimal solution -- dashed red.}
    \label{fig:control}
\end{figure*}

The first step to obtain the control policy in DPC is generating the synthetic dataset that samples the vector of control parameters $\xi$ with a probability distribution based on a priori knowledge. This dataset is used for training and development of the neural policy. As mentioned previously, the vector of control parameters contains the initial conditions of the states, $x_0$, and an $N$-steps-long window of future disturbances, $[d_0,\dots,d_{N-1}]$. Here, the reference is included as a constant term in the loss function and, thus, is not considered as a control parameter. The initial conditions of the state variable were sampled from a uniform probability distribution within the defined limits $x_{0|1}\!\sim\!\pazocal{U}(0,8.4)$ and $x_{0|2}\!\sim\!\pazocal{U}(0,3.6)$, covering the parametric space evenly. However, for disturbances, the synthetic data were generated on the basis of the available disturbance data, that is used for testing. For $d_1$, the available data appeared to resemble a Beta distribution with parameters $\alpha\!=\!0.6$ and $\beta\!=\!1.4$, thus $d_{0,\dots,N|1}\!\sim\!\text{Beta}(0.6,1.4)$ was generated and subsequently scaled with the factor of $7$. In the case of disturbance $d_2$, the data was generated using a custom algorithm that produces a signal with peaks. The amplitude of these peaks is randomly selected from the range $[1,16]$, and their duration varies between $[2,5]$ discrete time steps, replicating the testing data; see Fig. \ref{fig:control}.

In this numerical study, for all cases, the neural control policy was trained using a generated dataset consisting of $24000$ samples, while $4000$ samples were used for development, with a batch size of $n_\mathrm{b}\!=\!2000$. Algorithm~\ref{alg:training} outlines the MI-DPC training procedure of the neural control policy, where the Adam optimizer with a learning rate of $\zeta\!=\!3e^{-4}$ was used. Layer normalization was applied to each layer. In all instances, the policy was trained for up to $1000$ epochs, with early stopping being triggered after 80 consecutive bad counts. In the cases of DPC with Sigmoid and Softmax STE, all layers of the control policy were of size $120$, where the component $f^{(h)}_{\theta^{(h)}}$ was represented by a single layer and $f^{(u)}_{\theta^{(u)}}$ by two fully connected layers, each with hyperbolic tangent (tanh) activation. The neural component $f^{(\delta)}_{\theta^{(\delta)}}$ was also represented by two fully connected layers with scaled exponential linear unit (SELU) activation. For the learnable threshold method, the same architecture was used for both $\phi_{1|\theta_1}$ and $\phi_{2|\theta_2}$, where the layers were sized down to $95$, producing a neural model of a similar number of trainable parameters. In all cases, the scale coefficient was set to $\eta\!=\!10$ and the temperature to $\tau\!=\!0.5$. In the STE with Sigmoid, the relaxed solution of the integer control input was clipped using $\max\big(\!-\!0.5,\,\min(3.5,\,y^{(\delta_1)})\big)$, ensuring that the rounded values remain within the feasible integer set $\{0,1,2,3\}$.

In addition, a dropout regularization with a dropout probability of $0.1$ was applied only to the layers that compute the continuous control inputs because we observed an undesirable phenomenon when applying it to all layers. Although applying the dropout to all layers occasionally produced a controller with lower total loss, it also introduced a highly volatile behavior of the integer input. To be exact, during periods of zero disturbance $d_2$, for example from hours 20 to 30 in Fig.\ref{fig:control}, a high-frequency switching of the integer input occurred between the values $0$ and $1$, which caused slight oscillations of $x_2$ around the reference.

The DPC controllers and simulation results were obtained using a differentiable programming library NeuroMANCER v1.5.0 \cite{neuromancer2023}, running with PyTorch v2.5.0.  

\begin{algorithm}[t] \small
\caption{Training the neural policy: STE with Sigmoid}
\label{alg:training}
\begin{algorithmic}[1]
\Require Training dataset $\mathrm{D}_\xi$, loss function $\ell$, neural control policy $\pi_\theta$, learning rate $\zeta$, number of epochs $n_\text{epoch}$
\State Initialize parameters $\theta$
\Repeat
    %{\text{epoch} = 1 to $n_\text{epochs}$}
\For{$\xi$ in $\mathrm{D}_\xi$} \Comment{{\footnotesize {\color{gray} batch $\xi\in\mathbb{R}^{n_\mathrm{b}\times n_\xi}$}}}
    \State $\xi_0=[x_0^\mathsf{T},d_0^\mathsf{T},d_1^\mathsf{T},\dots,d_{N-1}^\mathsf{T}]$ \Comment{{{\footnotesize \color{gray}Initial vector}}}
        \State $\Gamma=[\;]$ \Comment{{\footnotesize{\color{gray} Initialize empty vector}}}
    \For{$k\!=\!0$ to $N\!-\!1$} \Comment{{\footnotesize{\color{gray}$N$ prediction horizon}}}
    \State Policy evaluation: $[y^{(u)}_{k},y^{(\delta)}_{k}]\!=\!\pi_\theta(\xi_{k})$ 
    \State Rounding: $\delta_k\!=\!\lfloor y^{(\delta)}_k\rceil$, $u_k\!=\!y^{(u)}_k$
    \State Dynamics: $x_{k+1}\!=\!Ax_k\!+\!B_uu_k\!+\!B_\delta\delta_k\!+\!Ed_k$ 
    \State Append: $\Gamma\leftarrow [x_k,u_k,\delta_k]$
    \State Iterate: $\xi_{k+1}\!=\![x_{k+1},d_{k+1},\dots,d_{N-1},\overset{\Lsh}{\bf 0}]$
    \Comment{\newline{\footnotesize {\color{gray} $\!\!\!\!\!\!\!\!$Sliding window: $[d_0,\dots,d_{N-1}]\!\xrightarrow{\text{slide}}\![d_1,\dots,d_{N-1},\overset{\Lsh}{\bf 0}]$ with 0 padding}}}
    \EndFor
    \State Calculate the gradients: $\nabla_\theta \ell(\Gamma)$
    % \State Calculate gradients: $\nabla_\theta L$ 
    \State Update the parameters: $\theta\leftarrow\theta-\zeta\frac{\uppartial \ell}{\uppartial \theta}$ 
    % \Comment{{\color{gray} }}
\EndFor
\Until{last epoch {\bf or} early stopping criteria}
    % \Return $w$
\end{algorithmic}
\end{algorithm}
\subsection{Simulation results}
In Tab.~\ref{tab:results}, based on the numerical experiments, we summarize the performance of MI-DPC using the three rounding strategies and compare them with the optimal solution of the hard-constrained MIQP formulation \eqref{eq:miqp} computed using CPLEX v12.10, for various lengths of $N$. The simulations were implemented in a receding horizon manner. For each case, the results were obtained by evaluating closed-loop simulations of \num{1872} steps, equivalent to \num{6.5} days, considering $20$ different initial conditions. In Fig. \ref{fig:control}, the first \num{3} days of the closed-loop experiment are shown, comparing MI-DPC with Sigmoid STE to optimal trajectories with a horizon of $20$, where $e_1$ and $e_2$ denote the difference between the optimal and MI-DPC state trajectories $x_1$ and $x_2$, respectively. The mean loss metric was computed as the arithmetic mean of the stepwise loss of the closed-loop data as
\begin{equation*}
    \ell_\mathrm{mean}=\frac{1}{n_\mathrm{ic} n_\mathrm{sim}}\sum_{i=1}^{n_\mathrm{ic}}\sum_{k=0}^{n_\mathrm{sim}} \ell_\pazocal{K}(x_{k|i},u_{k|i},\delta_{k|i}),
\end{equation*}
where $n_\mathrm{ic}$ and $n_\mathrm{sim}$ denote the number of initial conditions and the length of the simulation, respectively. Based on this metric, Tab.~\ref{tab:results} 
% and Fig. \ref{fig:loss}
indicates that the relative suboptimality margin (RSM) of the MI-DPC decreases as the length of the prediction horizon increases. This phenomenon is inherent to DPC and is caused by the fact that, within the training process, longer horizons allow backpropagation through longer sequences, capturing the transient system dynamics better, and thus approximating the solution to the parametric MI-OCP more accurately. Moreover, the mean inference time remains roughly constant due to the same network sizes and architectures used across different $N$, where a slight increase in the number of trainable parameters is present due to the varying size of the input layer. Compared to other methods, the LT provided the best overall performance, but was the slowest in inference and training despite the lower number of trainable parameters. This is due to the introduction of more floating-point operations by the correction layer. Over-all, all studied methods exhibit similar control performance. Notably, based on trial and error for this numerical example, the observation of STE with Sigmoid suggests that it is the least affected by varying hyperparameters; however, a com-prehensive ablation study would be necessary to confirm this.
\begin{figure}[tb]
    \begin{center}
        \begingroup
       \footnotesize
       \resizebox{\columnwidth}{!}{\input{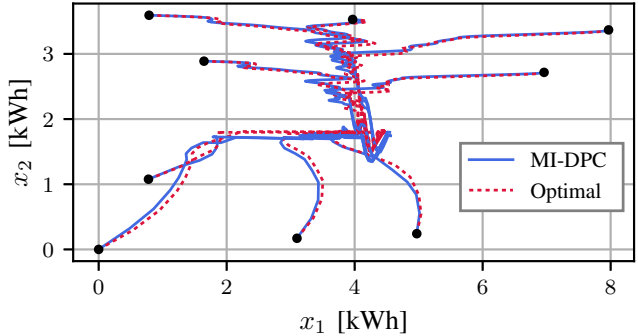}}
    \endgroup
    \end{center}
    \caption{Closed-loop state trajectories obtained with MI-DPC with Softmax STE compared with optimal state trajectories for $N\!=\!25$, simulated for $80$ time steps from different initial conditions marked by black dots.}\label{fig:phase_plot}
\end{figure}
\vspace{-2ex}
% But in case the feasible integers are not evenly spaced, for example, $\delta\!\in\!\{0,1,30\}$, only the Softmax method would be applicable.

Throughout the simulations performed, MI-DPC did not provide an infeasible solution with respect to the state and input constraints, although the feasibility of the solution is not guaranteed. On the other hand, CPLEX violated feasibility in terms of computation time, where a maximum of \SI{30}{\second} was considered to be an acceptable time to compute the control input. The frequency of occurrence of this violation was \SI{0.16}{\percent} for $N\!=25$ and \SI{3.62}{\percent} times for $N\!=30$ steps. %Out of \num{37440} total samples, this violation has occurred $60$ times for $N\!=\!25$ and \num{1356} times for $N\!=\!30$ steps.
% out of \num{37440} samples in total.
Compared with CPLEX, the proposed MI-DPC framework was shown to be very effective for longer horizons in terms of reducing the computation time and delivering performance with a low value of RSM. For example, for $N\!=30$, MI-DPC led to a reduction in mean inference time by four orders of magnitude, while operating with a relative suboptimality margin of \SI{1}{\percent}. This indicates that the proposed methodology may be suitable for embedded implementations or solving MI-OCPs even on edge devices with limited computational resources. Note that in Tab.~\ref{tab:results}, the comparison for $N\!=\!40$ was omitted, as the computation of the optimal solution with CPLEX was prohibitively time-consuming. Finally, Fig.~\ref{fig:phase_plot} shows the closed-loop state trajectories obtained with MI-DPC with Softmax STE and compares these with the optimal ones for different initial conditions, assuming $N\!=\!25$ steps.

\begin{table}[tb]
    \caption{Comparison of different approaches to solve the sample MI-OCP for various lengths of the prediction horizon.}
    \begin{threeparttable} % Begin threeparttable environment
    \centering
    \renewcommand{\arraystretch}{1.}
    \footnotesize
    \setlength{\tabcolsep}{3pt} % Reduce space between columns
    \begin{tabular}{cccccccc}
        \toprule
        &&  \multicolumn{6}{c}{Prediction horizon $N$}\\\cline{3-8} \rule{0pt}{3ex}
        Approach & Metric & \textbf{$10$} & \textbf{$15$} & \textbf{$20$} & \textbf{$25$} & \textbf{$30$} & \textbf{$40$} \\
        \midrule
        \multirow{5}{*}{\shortstack{MI-DPC\\with\\Sigmoid\\STE}} 
        & $\ell_\mathrm{mean}$  & 6.7101 & 4.7318 & 4.0879 &  3.9581 & 3.8702 & 3.8436 \\
        & RSM (\%) & 15.13\% & 7.25\% & 1.81\% & 1.77\% & 0.60\% & -- \\
        & MIT (s)\tnote{\,1} & 0.0004 & 0.0004 & 0.0004 & 0.0004 & 0.0004 & 0.0004 \\
        & NTP (-) & 82603 & 84003 & 85403 & 86803 & 88203 & 91003 \\
        
        & TT (s)\tnote{\,2} & 585.4 & 664.9 & 721.9 & 789.2 & 862.4 & 868.6 \\
        % & Epochs & 760 & 739 & 1000 & 758 & 1000 & 827 \\
        \midrule
        \multirow{5}{*}{\shortstack{MI-DPC\\with\\Softmax\\STE}} 
        & $\ell_\mathrm{mean}$ & 6.7930 & 4.7851 & 4.1327 & 3.9483 & 3.8721 & 3.8656 \\
        & RSM (\%) &16.55\%& 8.45\% & 2.93\% & 1.53\% & 0.65\% & --  \\
        & MIT (s)\tnote{\,1}  & 0.0004 & 0.0005 & 0.0005 & 0.0005 & 0.0005 & 0.0005 \\
        & NTP (-) & 83026 & 84426 & 85826 & 87226 & 88626 & 91426 \\
        
        & TT (s)\tnote{\,2}  & 419.3 & 557.8 & 734.8 & 589.9 & 628.4 & 621.5 \\
        % & Epochs & 868 & 680 & 917 & 693 & 1000 & 1000 \\
        \midrule
        \multirow{5}{*}{\shortstack{MI-DPC\\with\\Learnable\\Threshold}} 
        &  $\ell_\mathrm{mean}$ & 6.4613 & 4.5480 & 4.0909 & 3.9153 & 3.8753 & 3.8443 \\
        & RSM (\%) &10.86\% &3.08\% &1.88\% &0.68\% &0.73\% & --\\
        & MIT (s)\tnote{\,1}  & 0.0010 & 0.0010 & 0.0010 & 0.0010 & 0.0010 & 0.0010 \\
        & NTP (-) & 78191 & 80091 & 81991 & 83891 & 85791 & 89591 \\
        & TT (s)\tnote{\,2} & 700.2 & 926.4 & 1110.8 & 1293.7 & 1282.7 & 1659.4 \\
        % & Epochs & 100 & 100 & 100 & 100 & 100 & 100 \\
        \midrule
        \multirow{3}{*}{\shortstack{Optimal\\(MIQP\tnote{\,3}$\;\;$)}} 
        &  $\ell_\mathrm{mean}$ & 5.8287 & 4.4124 & 4.0157 & 3.8893 & 3.8474 & \dag \\
        & MIT (s)\tnote{\,1}   & 0.0041 & 0.015 & 0.0913 & 0.6071 & 3.9404 & \dag \\
        & FUP (\%) & 0\% & 0\% & 0\% & 0.16\% & 3.62\% & \dag \\
        \bottomrule
    
    \end{tabular}
\begin{tablenotes}
\footnotesize
\item Used acronyms: RSM -- Relative Suboptimality Margin, MIT -- Mean Inference Time, NTP -- Number of Trainable Parameters, TT -- Training Time,  FUP -- Fraction of Unsolved Problems within the \SI{30}{\second} time limit.
\item[1] Measured using a single thread of Intel Xeon Gold 6226R.
\item[2] Trained using an NVIDIA L40s.
\item[3] Solved with the CPLEX branch-and-bound algorithm.
% \item[4] Relative suboptimality margin calculated as.% $\left(\frac{\ell_\mathrm{DPC}}{\ell_\mathrm{optimal}}-1\right) 100\%$.
\item[\dag] Failed to perform simulations within \num{40} hours.
\end{tablenotes}
\end{threeparttable}
\label{tab:results}
\end{table}
% \vspace{-0.1cm}
% \begin{figure}[tbh]
%     \begin{center}
%         \begingroup
%        \footnotesize
%        \resizebox{\columnwidth}{!}{\input{figures/loss_plot.pgf}}
%     \endgroup
%     \end{center}
%     \caption{Comparison of mean loss for different lengths of $N$.}
%     \label{fig:loss}
% \end{figure}

% \newpage
% ~
% \vfill

%%%%%%%%%%%%%%%%%%%%%%%%%%%%%%%%%%%%%%%%%%%%%%%%%%%%%%%%%%%%%%%%%%%%%%%%%%%%%%%%
\section{CONCLUSION}
\label{sec:5}
This work introduced a scalable and computationally efficient approach to solving parametric mixed-integer optimal control problems by leveraging the established framework of differentiable predictive control. The use of differentiable surrogates enables approximate gradient-based optimization of the neural control policy, while integrality of the discrete control inputs is enforced through non-differentiable operations such as rounding. The results of the closed-loop simulation experiments indicate that the proposed self-supervised learning methodology can lead to near-optimal control performance while reducing computational time by several orders of magnitude, potentially allowing for implementations on computationally constrained embedded hardware. Importantly, the proposed approach does not guarantee closed-loop stability, global optimality, or feasibility. In line with this, our future work will focus on deriving rigorous stability and feasibility guarantees. The proposed approach can also be leveraged by employing pretrained neural policies to warm start the optimization solvers. In future work, we also plan to examine the performance of the proposed method on larger systems with nonlinear dynamics and constraints. 
% \vspace{-0.22cm}
% \newpage

%%%%%%%%%%%%%%%%%%%%%%%%%%%%%%%%%%%%%%%%%%%%%%%%%%%%%%%%%%%%%%%%%%%%%%%%%%%%%%%%
% \section{ACKNOWLEDGMENTS}

% The authors gratefully acknowledge the contribution of National Research Organization and reviewers' comments.

%%%%%%%%%%%%%%%%%%%%%%%%%%%%%%%%%%%%%%%%%%%%%%%%%%%%%%%%%%%%%%%%%%%%%%%%%%%%%%%%

\setstretch{0.965}

\bibliography{bibliography}
\bibliographystyle{IEEEtran}

% \begin{thebibliography}{99}

% \bibitem{c1}
% J.G.F. Francis, The QR Transformation I, {\it Comput. J.}, vol. 4, 1961, pp 265-271.

% \bibitem{c2}
% H. Kwakernaak and R. Sivan, {\it Modern Signals and Systems}, Prentice Hall, Englewood Cliffs, NJ; 1991.

% \bibitem{c3}
% D. Boley and R. Maier, "A Parallel QR Algorithm for the Non-Symmetric Eigenvalue Algorithm", {\it in Third SIAM Conference on Applied Linear Algebra}, Madison, WI, 1988, pp. A20.

% \end{thebibliography}

\end{document}